\documentclass[twoside,slac_one]{revtex4}
\usepackage{graphicx}
\usepackage{fancyhdr}
\usepackage{amsmath} 
\usepackage{bm}
\usepackage{amsxtra}
\usepackage{amssymb}
\usepackage{amsthm}
\usepackage{latexsym}
\usepackage{lscape}
\usepackage{atlasphysics} 
\usepackage{booktabs} 
\usepackage{subfigure}

\begin{document}

\title{Measurement of the \ZZ production cross section and limits on anomalous neutral triple gauge couplings in proton-proton collisions at $\mathbf{\sqrt{s} =7}\TeV$ with the ATLAS detector}

%

\author{Shih-Chieh Hsu on behalf of ATLAS}
\affiliation{Lawrence Berkeley National Lab, Berkeley, CA, USA}

\begin{abstract}
A measurement of the \ZZ production cross section in proton-proton collisions
at $\sqrt{s} = 7\TeV$ using data collected by the ATLAS experiment
at the LHC is presented.
In a data sample corresponding to an integrated luminosity of $1.02~\ifb$~\cite{ATLAS-lumi2011},
12 events containing two Z boson candidates decaying to electrons and/or muons
were observed.
The expected background contribution is
$0.3^{+0.9}_{-0.3} \text{(stat)} ^{+0.4}_{-0.3} \text{(syst)}$ events.
The total cross section for on-shell \ZZ production has been determined to be
$\sigma_{ZZ}^\mathrm{tot}= 8.4^{+2.7}_{-2.3}\text{ (stat)}^{+0.4}_{-0.7} \text{ (syst) }\pm 0.3\textrm{ (lumi) }\pb$ and is compatible with the Standard Model
expectation of $6.5^{+0.3}_{-0.2}~\pb$ calculated at the next-to-leading order in QCD.
Limits on anomalous neutral triple gauge boson couplings are derived.
\end{abstract}

\maketitle

\thispagestyle{fancy}


\section{Introduction}\label{sec:Introduction}
\begin{figure}[htbp]
  \begin{center}
  \includegraphics[width=0.7\textwidth]{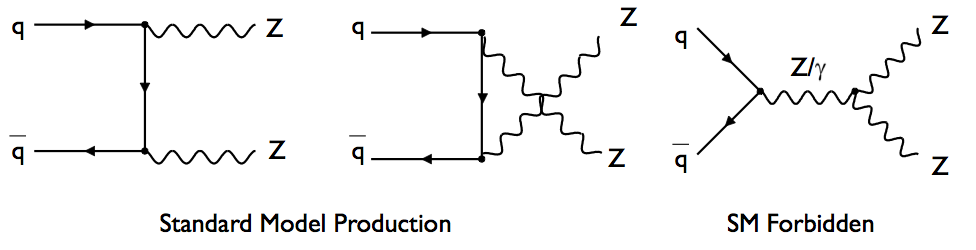}
  \caption{Tree-level Feynman diagrams for \ZZ production through the \qqbar\ initial state in 
           hadron colliders. The $s$-channel diagram, on the right, contains the $ZZZ$ 
           ($ZZ\gamma$) neutral triple gauge boson coupling vertex, which is zero in the Standard 
           Model.}
\label{fig:LOdiagrams}
\end{center}
\end{figure}

The production of pairs of \Z\ bosons at the LHC is of great interest since 
it provides a unique opportunity to test the
predictions of the electroweak sector of the Standard Model 
at the TeV energy scale, and it is the irreducible background to the 
search for the Higgs boson in the $\Hboson\ra ZZ$ decay channel
or new phenomena beyond the SM~\cite{zzres1,zzres2}.
In the Standard Model, \ZZ\ production proceeds via quark-antiquark $t$-channel 
annihilation; 
Figure~\ref{fig:LOdiagrams} shows the leading-order Feynman diagrams for \ZZ production 
from \qqbar\ initial states.
The $ZZZ$ and $ZZ\gamma$ neutral triple gauge boson couplings (nTGCs) are zero in the Standard Model, hence
there is no contribution from $s$-channel \qqbar\ annihilation at tree 
level. At the one-loop level the contribution is $\mathcal{O}(10^{-4})$~\cite{Gounaris:2000dn}.
Many models of physics beyond the Standard Model
predict values of these couplings at the level of $10^{-4}$ to $10^{-3}$~\cite{Ellison:1998}. 
Most non-zero values of $ZZZ$ and $ZZ\gamma$ couplings  
increase the \ZZ cross section at high \ZZ invariant mass and high transverse
momentum of the \Z\ bosons~\cite{Baur:2000ae}. 
\ZZ production has been studied in \epem\ collisions at 
LEP~\cite{Barate:1999jj,Abdallah:2003dv,Acciarri:1999ug,Abbiendi:2003va,bib:LEPEW2006}
and in antiproton-proton collisions at the Tevatron~\cite{bib:CDF_ZZ,bib:D0_ZZ1,bib:D0_ZZ2,bib:D0_ZZ3}.
No deviation of the measured cross section from Standard Model expectation has been observed, 
allowing limits on anomalous nTGCs to be set~\cite{bib:LEPEW2006,bib:D0_ZZ1}. 

\vspace{-0.5cm}

This paper presents the first measurement of $\ZZ$\footnote{Throughout this paper \Z\ should be taken 
to mean $\Z/\gamma^{*}$.} production in proton-proton
collisions at a centre-of-mass energy $\sqrt{s}$ of 7 \TeV, and limits on the anomalous nTGCs~\cite{ATLAS-CONF-2011-107}. The measurement uses a data sample
corresponding to an integrated luminosity of 1.02 \ifb\ collected by the ATLAS detector at 
the LHC between February and June 2011. The cross section for on-shell \ZZ production is
predicted at next-to-leading order in QCD (NLO) to be $6.5^{+0.3}_{-0.2}\ \pb$;
this includes a $\sim$6\% contribution from gluon fusion~\cite{Campbell:2011}.   
Candidate \ZZ events are reconstructed in
the $\ZZ\rightarrow\ll\ll$ decay channel, where $\ell$ can be an electron or muon. 
Although this channel constitutes only $\sim$0.5\% of the total on-shell \ZZ cross section, 
the expected signal to background ratio of four high transverse-momentum and isolated leptons is $\sim$30.  

To reduce systematic uncertainties
the cross section is measured
within a phase-space that corresponds closely to the experimental selection cuts, namely
requiring the mass of both lepton pairs to be between 66 \GeV\ and 116 \GeV\ and all
four leptons to be within the pseudorapidity\footnote{ATLAS uses a right-handed coordinate
system with its origin at the nominal interaction point in the centre of the detector and
the $z$-axis along the beam pipe. The $x$-axis points from the interaction point to
the centre of the LHC ring, and the $y$-axis points upwards. Cylindrical coordinates ($r$,$\phi$)
are used in the transverse plane, $\phi$ being the azimuthal angle around the beam pipe.
The pseudorapidity $\eta$ is defined in terms of the polar angle $\theta$ as
$\eta = - \ln\tan(\theta/2)$.} range $|\eta| < 2.5$ and have transverse
momentum (\pT) greater than 15 \GeV. This is termed the `fiducial' cross section.
The total \ZZ cross section in the on-shell approximation
is obtained from the fiducial cross section
using the known $\Z\rightarrow\ll$ branching ratio and a correction factor
for the kinematic and geometrical acceptance.

Anomalous nTGCs for on-shell \ZZ production can be parameterized by two CP-violating ($f_4^V$) and two 
CP-conserving ($f_5^V$) complex parameters ($V = \Z, \gamma$) which are zero in the Standard 
Model~\cite{Baur:2000ae}.
To ensure the partial-wave unitarity, a form-factor parameterization is introduced to cause
the couplings to vanish at high center-of-mass energy  $\sqrt{\hat{s}}$: 
$f_i^V = f_{i0}^{V}/(1+ \hat{s}/\Lambda^{2})^{n}$.
Here, $\Lambda$ is the energy scale at which physics beyond the Standard Model will be directly observable, $f^{V}_{i0}$ are the
low-energy approximations of the couplings, and $n$ is the form-factor
power. Following Ref.~\cite{Baur:2000ae}, we set $n = 3$
and the form-factor scale $\Lambda$ to 2 \TeV, so that expected limits
are within the values provided by unitarity at LHC energies. The results with energy cutoff $\Lambda=\infty$ are also 
presented as a comparison in the unitarity violation scheme.


\section{Detector, Data and Monte Carlo Simulation}\label{sec:Data}
The ATLAS detector~\cite{bib:ATLASDetectorPaper} consists of an inner tracking detector 
surrounded by a superconducting solenoid, electromagnetic and
hadronic calorimeters and a muon spectrometer with a toroidal magnetic field. The inner
detector, in combination with the 2\,T magnetic field from the solenoid, provides precision tracking of charged particles for $|\eta|<2.5$. It consists of a
silicon pixel detector, a silicon strip detector and a straw tube tracker that also produces
transition radiation measurements for electron identification. The calorimeter system  covers the pseudorapidity range $|\eta| < 4.9$. 
It is composed of sampling calorimeters with either liquid argon (LAr) or scintillating tiles as the active media. 
In the region $|\eta|<2.5$ the electromagnetic LAr
calorimeter is finely segmented and plays an important role in electron identification.
The muon spectrometer has separate trigger and high-precision tracking chambers which provide muon
identification in $|\eta|<2.7$. 
A three-level trigger system selects events to record for offline
analysis. The events in this analysis were selected with  single-lepton triggers with transverse
momentum thresholds of 20 \GeV\ for electrons and 18 \GeV\ for muons. 

This measurement uses a data sample of proton-proton collisions at $\sqrt{s} = 7 \TeV$
recorded between February and June 2011. Data periods flagged with data quality problems
that affect the lepton reconstruction are removed. After data quality
cuts, the total integrated luminosity used in the analysis is \mbox{1.02 \ifb.} The 
luminosity uncertainty is 3.7\%~\cite{ATLAS-lumi2011}.

The signal acceptance is determined from a detailed Monte Carlo simulation. 
The leading order (LO) generator \pythia~\cite{pythia} with the MRST modified
LO Parton Density Function (PDF) set~\cite{bib:MRST2007LOmod} 
is used to model  $pp \ra \ZZ \ra \ll\ll$ events,
where $\ell$ includes electrons, muons and tau leptons. 
The \pythia simulation includes the interference terms between the \Z\ and $\gamma^{*}$ diagrams. 
The minimum invariant mass generated for each $\Z/\gamma$ boson is set to \mbox{12 \GeV}. When calculating the expected 
number of signal events, the
predictions of \pythia are normalized to the NLO calculation of MCFM~\cite{Campbell:2011} using
the MSTW2008~\cite{bib:MSTW2008} NLO PDF set. The normalization factor, calculated within the 
phase-space of our fiducial cross section measurement, is 1.44.

Background processes are estimated from the data and are cross-checked with predictions from Monte
Carlo simulations.
\mcatnlo~\cite{bib:mcatnlo} is used to model the diboson processes \WW and \WZ, \ttbar\ and single 
top-quark events, and \madgraph~\cite{madgraph} is used for the $W/Z+\gamma$ final state. 
\W\ or \Z\ gauge bosons produced in association with jets are modelled with \alpgen~\cite{alpgen},
except for those with a $\tau$ lepton in the final state where \pythia was used.
Events with dileptons from Drell-Yan ($10 \GeV < m_{\ll} <40 \GeV $) production are also modelled
with \alpgen or \pythia, depending on the presence of $\tau$ leptons in the final state.
Events with heavy flavour dijets are modelled with \pythiaB~\cite{pythiab}.
 
Detector response was simulated~\cite{bib:ATLASMCpaper} with a program based on GEANT4~\cite{bib:geant4a}.
The luminosity in a single bunch-crossing is high enough to produce several $pp$ collisions per
bunch crossing, the mean number typically being 6--8. Additional inelastic $pp$ events are
included in the simulation, distributed so as to reproduce the expected number of collisions per 
bunch-crossing in the data. 

\section{Event selection, signal acceptance and efficiency}\label{sec:EventSelection}
 \begin{figure}[htbp]
 \begin{center}
  \includegraphics[width=0.5\textwidth]{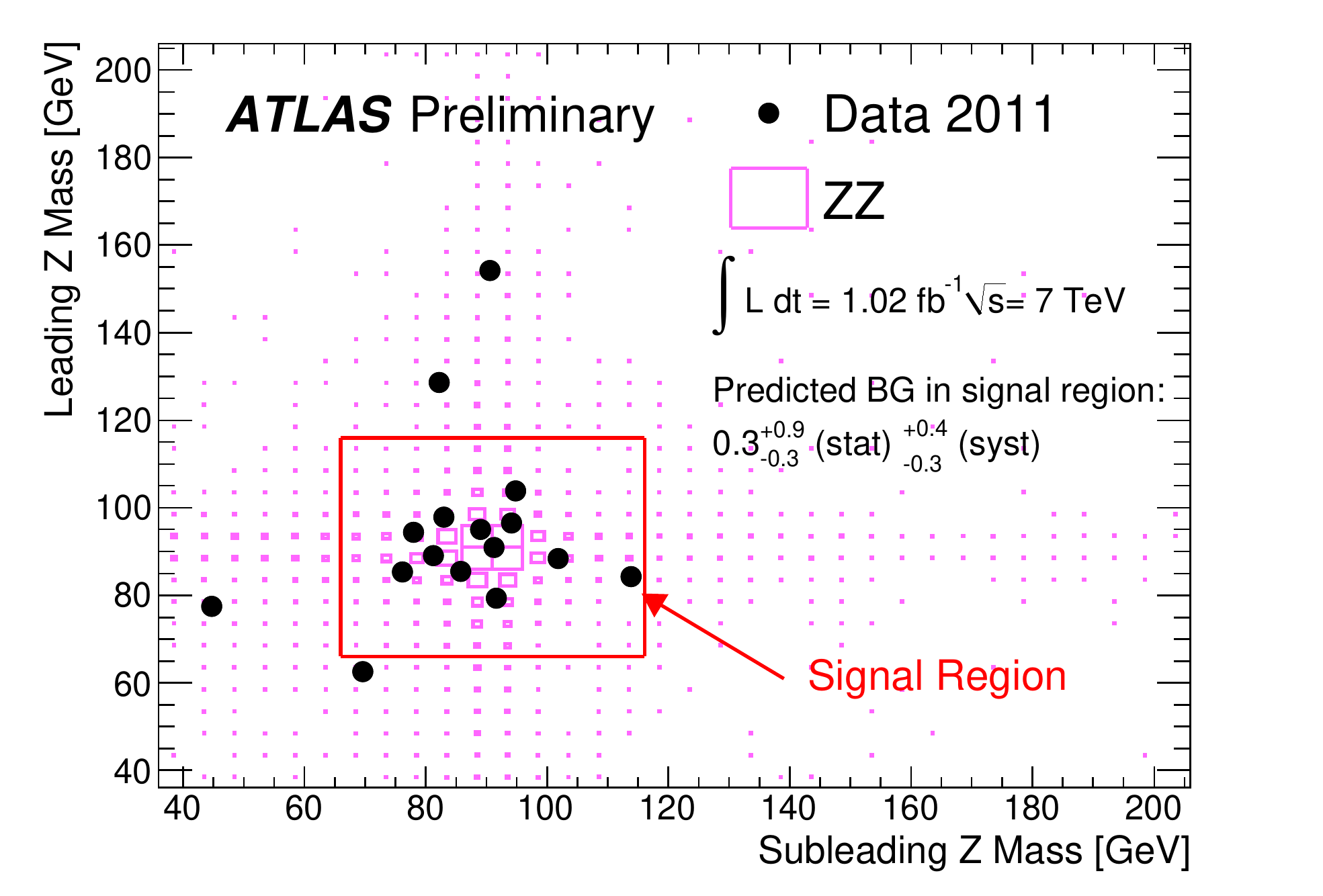}\hfill
  \caption{\small The mass of the leading (higher lepton pair $\pt$) \Z\ candidate  versus the mass of the subleading (lower lepton pair $\pt$) \Z\ candidate.
  The events observed in the data are shown as solid circles and the signal prediction from simulation as pink boxes.
  The red box indicates the signal region defined by the cuts on the \Z\ candidate masses.
   }
 \label{fig:mass2d}
 \end{center}
 \end{figure}
In order to remove non-collision background, events are required to contain at least one vertex
formed from at least three good tracks.
The vertex with the largest sum of the $\pT^2$ computed with the associated tracks is selected as 
the primary vertex.  
Signal events are characterized by four high-\pT, isolated electrons or muons, 
in three channels: \ee\ee, \mumu\mumu and \ee\mumu. Lepton candidates are required to be
consistent with originating from the primary vertex.

Electrons are reconstructed from a cluster in the electromagnetic calorimeter matched to a track in the 
inner detector~\cite{bib:ATLAS_W_Z}. 
Electron candidates are required to have a transverse energy (measured in the calorimeter)
of at least 15 \GeV\ and a pseudorapidity of $|\eta| < 2.47$. They must be isolated, using the same 
criterion as for muons, calculating the $\Sigma p_{T}$ around the electron track.
Electron candidates within $\Delta R = 0.1$ of any selected muon
are rejected, and if two electron candidates are within $\Delta R = 0.1$ of each other the one with
the lower \pT\ is rejected.
The overall reconstruction and identification efficiency varies as a function of \pT\ from
63\% at 15 \GeV\ to 81\% at 45 \GeV.

Muons are identified by matching tracks (or track segments) reconstructed in the muon spectrometer 
to tracks reconstructed in the inner detector~\cite{bib:ATLAS_W_Z}. 
Their momentum is calculated by combining the information from the two subsystems
and correcting for the energy lost in the calorimeter. Only muons with $\pT >15\GeV$
and $|\eta| < 2.5$ are considered. 
In order to reject muons from the decay of heavy quarks, isolated muons are selected,
by requiring
the scalar sum of the transverse momenta ($\Sigma \pT$) of other tracks inside a cone of 
$\Delta R \equiv \sqrt{\Delta \phi^2 + \Delta \eta^2} = 0.2$ 
around the muon to be no more than 15\% of the muon $p_T$. 
The muon reconstruction and isolation efficiencies were measured in data, using 
a tag-and-probe technique on a large sample of $\Z\rightarrow\mu^+\mu^-$ events.
The overall reconstruction and identification efficiency varies as a function of \pT\ from
92\% at 15 \GeV\ to 95\% at 45 \GeV.

Selected events are required to have exactly four leptons selected as above, and to  have passed a single-muon or single-electron trigger. 
To ensure event selection at trigger efficiency plateau, at
least one of these leptons must have $\pT >$ 20 \GeV (25 \GeV) for a muon (electron) and match
to an object of the same flavor reconstructed online by the trigger system within $\Delta R <$ 0.1 (0.15).

Same-flavour, oppositely-charged lepton pairs are combined to form \Z\ candidates.
An event must contain two such pairs. 
In the \ee\ee and \mumu\mumu channels, ambiguities are resolved by choosing the pairing which
results in the smaller value of the sum of the two $|m_{\ll} - \mZ|$ values.
Figure~\ref{fig:mass2d} shows a scatter plot of the invariant mass 
of the leading (higher lepton pair \pT) lepton pair against that of the subleading (lower lepton pair \pT) lepton pair.
The events cluster in the region where both masses are around \mZ, with some contribution from events
with one \Z\ boson off-shell.
Events are required to contain two \Z\ candidates with invariant masses satisfying 
$66 \GeV < m_{\ll} < 116 \GeV$.

The reconstruction efficiency for the \ZZ candidates, including the
trigger efficiency, the lepton identification and reconstruction
efficiencies is derived from simulation. It is corrected with scale factors to account for
small differences in efficiencies between data and simulation.

The efficiencies of the single-lepton triggers have been determined as a function of lepton \pT\ using 
large samples of single $\Z\ra\ll$ events. 
The trigger efficiencies for the four-lepton events passing the
offline selections are obtained from simulation,
corrected by scale factors derived from the comparison between data and Monte Carlo of the single lepton
trigger efficiencies, and are found to be close to 100\% with uncertainty 0.04\%.

For lepton reconstruction and identification, the scale factors 
vary from unity by 1\%--13\% for electrons and 0.1\%--2\% for muons~\cite{ATLAS-CONF-2011-063} 
depending on the \pT\. The larger
discrepancies seen for electrons affect only the low-\pT\ region, and are due to mis-modelling of 
lateral shower shapes in simulation.
Systematic uncertainties on these scale factors are derived from efficiency measurements in the data.
A $\sim$1\% correction is applied to the calorimeter energy scale and resolution for electron
so that the $\Z\ra\ee$ invariant mass distribution in data is correctly reproduced by the simulation; similiary, a small correction (1\% at 15$\GeV$--2\% at 50$\GeV$) is applied to muon $\pt$.

The performance of the Monte Carlo simulation relative to the data has been checked using single \Z\ boson 
candidates. Same-flavour opposite-sign lepton pairs were selected using the criteria above, including the 
trigger requirement, and compared to simulation after applying all corrections. The invariant mass
distributions of dimuon and dielectron pairs are shown in Fig.~\ref{fig:zmm_zee_mass_DataMC}(a) and
(b) respectively; reasonable agreement is seen.
\vspace{-0.3cm}
 \begin{figure}[!htbp]
 \begin{center}
 \subfigure[]{
 \includegraphics[width=0.4\textwidth]{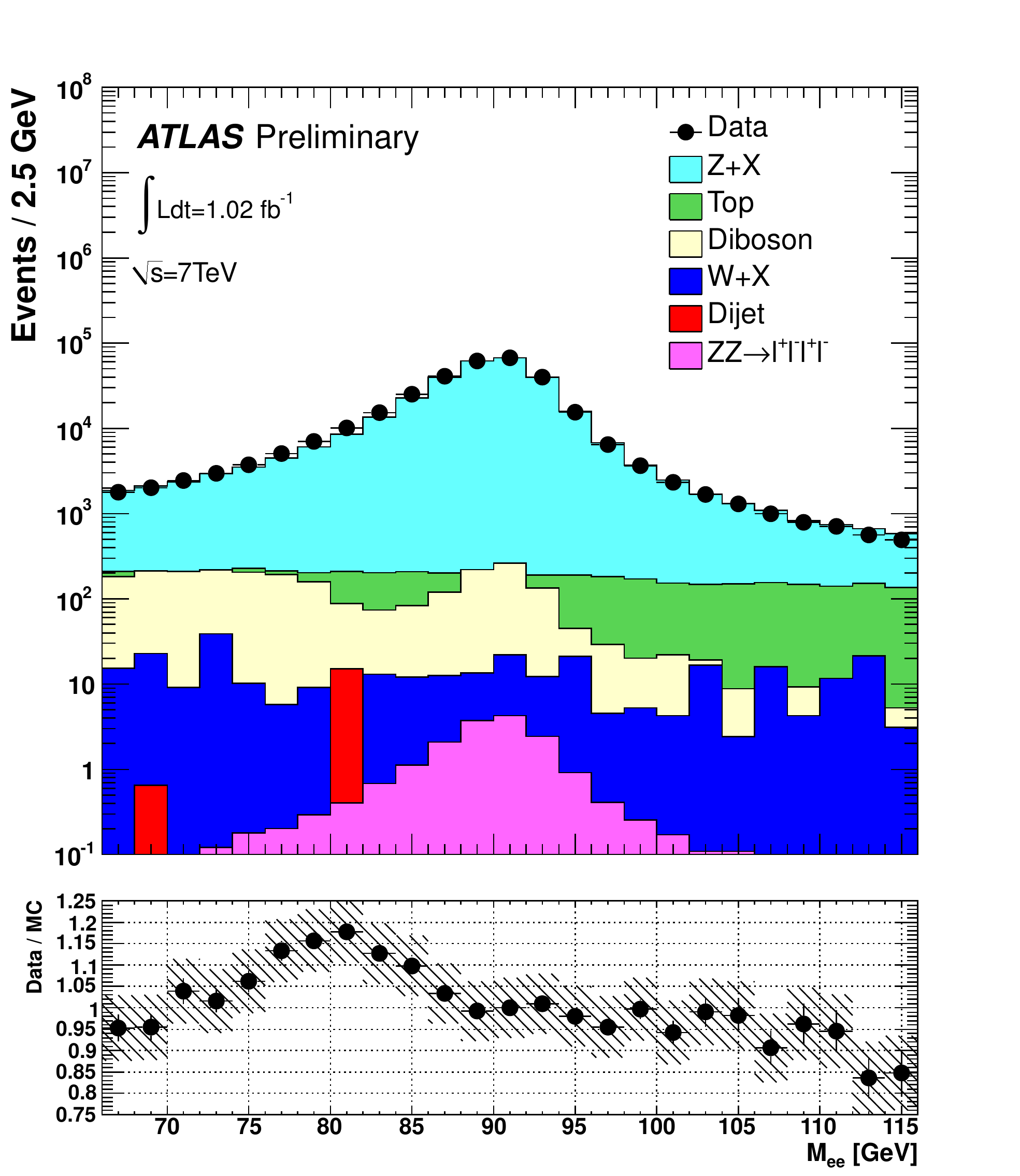} \hfill
 }
 \subfigure[]{
 \includegraphics[width=0.4\textwidth]{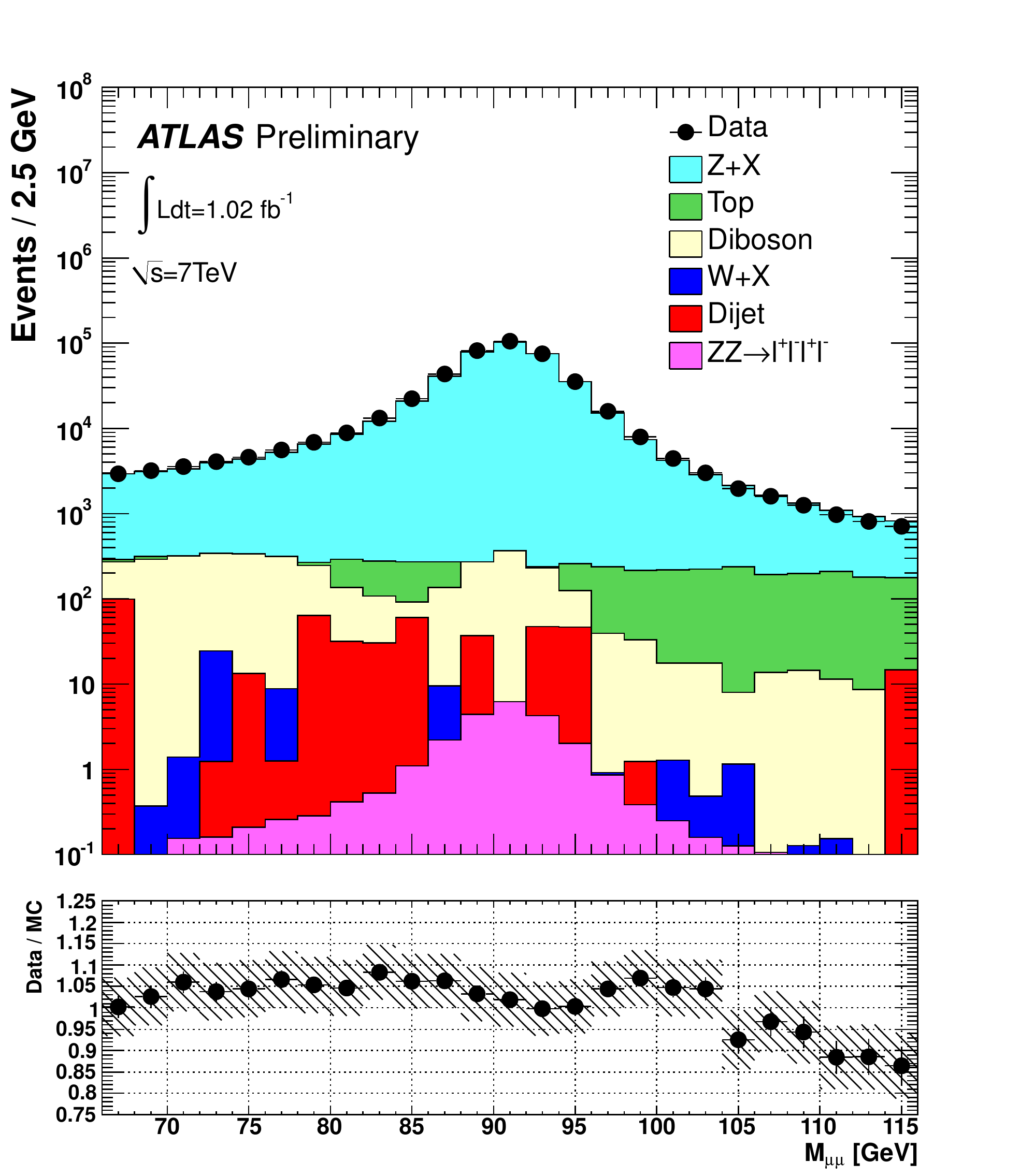} \hfill
 }
 \caption{\small Invariant mass of (a) dielectron \Z\ candidates and (b) dimuon \Z\ candidates.
          The different background sources are shown in different colours. The ratio between data 
          and Monte Carlo is shown in the lower part of each histogram; the total error (statistical
          plus systematic) is indicated by the shaded band. Diboson includes $WW$, $WZ$, $W\gamma$ and $Z\gamma$.
   }
 \label{fig:zmm_zee_mass_DataMC}
 \end{center}
 \end{figure}

\vspace{-0.6cm}
The overall efficiencies of the reconstruction and selection
criteria for events
generated within the fiducial phase-space are 41$\pm$3\%, 81$\pm$2\%, 57$\pm$2\% and 59$\pm$2\% for \ee\ee, \mumu\mumu, \ee\mumu\ and \ll\ll respectively. 
They include contributions of 1.6\% from $\ZZ\ra\ll\ll$ events generated
outside the fiducial phase-space and 0.3\% from events where one of the \Z\ bosons decays to tau leptons.
The lower signal expectation in the \ee\ee channel compared with the \mumu\mumu\
channel reflects the lower electron identification efficiency. 
The dominant systematic uncertainties arise from electron identification
(6.6\% in the \ee\ee final state, 3.1\% in the \ee\mumu final state) and 
from the muon reconstruction efficiency (2.0\% in \mumu\mumu and 1.0\% in \ee\mumu). 

\section{Background Estimation}\label{sec:Background}
Background to the \ZZ signal originates from events with a \Z\ (or \W) boson decaying to leptons
plus additional jets or photons ($W/Z+X$) and from top-quark pair-production ($t\bar{t}$) or single-top
production. 
The jets may be misidentified as electrons or contain electrons or muons from in-flight decays of
 pions, kaons, or heavy-flavoured hadrons; photons may be misidentified as electrons.
The majority of these background leptons are rejected by the isolation requirement, 
but some may satisfy the isolation cuts. 
Since Monte Carlo simulations may not model well the jet fragmentation in the tails of the isolation 
distributions, the background is estimated directly from the data.

To estimate the background contribution from four-lepton events in which one lepton does not
originate from the decay of a \Z\ boson but from a jet, a sample of events in the data containing
three leptons passing all selection criteria plus one `lepton-like jet' is identified;
such events are denoted 
$\ell \ell \ell j$.
For muons, the lepton-like
jets are muon candidates that fail the isolation requirement. For electrons, the lepton-like
jets are clusters in the electromagnetic calorimeter matched to inner
detector tracks that fail either or both of 
the full electron selection and 
the isolation requirement. 
The events
are otherwise required to pass the full event selection, treating the lepton-like jet as if it
were a fully identified lepton. This event sample is dominated by $\Z + X$ events. 
The background is then estimated by scaling this control sample
by a measured factor $f$ (\eta\ and \pT\ dependent, treated as uncorrelated 
in the two variables) 
which is the ratio of the probability for a jet to 
satisfy the full lepton criteria to the probability to satisfy the lepton-like jet criteria.
The background in which two selected leptons originate from a jet is treated similarly, by 
identifying a data sample with two leptons and two lepton-like jets;
such events are denoted 
$\ell \ell j j$.
To avoid double counting in the background estimate,
and to take account of the expected \ZZ contribution, 
the total number of background events is calculated as:
\begin{equation}
N(\text{background}) = N(\ell\ell\ell j)\times f - N(\ell\ell jj)\times f^2 - N(\ZZ\mbox{ in control region}).
\end{equation}
The factor $f$ is measured in a sample of data selected with single-lepton triggers with cuts
applied to suppress isolated leptons from \W\ and \Z\ bosons,
and corrected for the remaining small contribution of true leptons using simulation. 
A similar analysis is performed on Monte Carlo simulation; the larger of the statistical error on
$f$ determined from the data and the difference between data and simulation is taken as the 
systematic uncertainty in each \pT\ (or $\eta$) bin. 
This results in an average of systematic uncertainty of $\sim$30\% 
for each \pT (\eta)  bin except the lowest \pT\ (15 -- 20 \GeV) bin, for which there is a 
100\% systematic uncertainty.

\section{Cross section measurement}\label{sec:Results}
The numbers of expected and observed events after applying all selection cuts
are shown in Table~\ref{ta:selected_data_MC} and depicted in Figure~\ref{fig:kindists}.
The expected yields are compatible to theoretical predictions.
We observe 12 \ZZ candidates in data with a background expectation of 
$0.3^{+0.9}_{-0.3}$,
corresponding to a p-value of $3.4\times 10^{-6}$ equivalent to a one-sided Gaussian significance of 4.5$\sigma$. 
In the four-muon channel 8 events are
observed where only 
$3.3^{+1.0}_{-0.5}$
total signal plus background events are expected. The probability of the expected number
fluctuating up to 8 or more is 6.4\%.  

\begin{table}[htbp]
  \centering
  \begin{tabular}{lcccc} 
    \hline\hline
     Final State    & \ee\ee & \mumu\mumu & \ee\mumu & \ll\ll \\
    \midrule
    Observed        & 2 & 8 & 2 & 12 \\
    \midrule
     Bkg(data-driven)  &  0.01$^{+0.03 +0.05}_{-0.01 - 0.01}$ &$0.3^{+0.9}_{-0.3}\pm0.3$  & $<0.01^{+0.03}_{-0.01}$ &$0.3^{+0.9 +0.4}_{-0.3 -0.3}$\\ 
     \midrule
     Expected \ZZ    & 1.57$\pm$0.03$\pm$0.11   & 3.09$\pm$0.04$\pm$0.06 & 4.5$\pm$0.1$\pm$0.2 & 9.1$\pm$0.1$\pm$0.3 \\
    \hline
  \end{tabular}
\caption{Summary of observed events, total background contributions and expected signal
           in the individual four-lepton and combined channels. 
           The first error is statistical while the second is systematic. The uncertainty on the
           luminosity is not included. The errors on the background estimates span the 68\%
           confidence interval, which is not symmetric about the best estimate because the background
           cannot be negative.
}
  \label{ta:selected_data_MC}
\end{table}

The \ZZ fiducial cross section was determined by maximizing a profile likelihood. The profile likelihood is derived by maximizing a likelihood including systematic uncertainties as nuisance parameters with respect to those nuisance parameters for each cross section value.
The measured fiducial cross section
is:
\begin{eqnarray*}
\sigma_{ZZ\to \ll\ll}^\mathrm{fid} &=& 19^{+6}_{-5}\textrm{ (stat) } ^{+1}_{-2}\textrm{ (syst) }\pm 1 \textrm{ (lumi) }\mathrm{fb}
\end{eqnarray*}
where $\ll\ll$ refers to the sum of the \ee\ee, \ee\mumu\ and \mumu\ee\ final states.
The total cross section was determined similarly, but correcting for the known $\Z\to\ll$ branching
ratio and the acceptance of the fiducial cuts. The acceptance of the fiducial cuts, calculated at NLO 
using the program MCFM~\cite{Campbell:2011} version 6.0 with the MSTW2008 PDF set, is $0.507\pm0.008$ 
where the error arises from PDF uncertainties.
The measured value of the total on-shell \ZZ cross section is: 
\begin{eqnarray*}
\sigma_{ZZ}^\mathrm{tot} &=& 8.4^{+2.7}_{-2.3}\textrm{ (stat) }^{+0.4}_{-0.7}\textrm{ (syst) }\pm 0.3\textrm{ (lumi) }\mathrm{pb}.
\end{eqnarray*}
The result is statistically consistent with the NLO Standard Model total cross section for this
process of $6.5^{+0.3}_{-0.2}\ \pb$, calculated with MCFM and parton density function set MSTW2008.

\begin{figure}[htbp]
 \begin{center}
 \includegraphics[width=0.4\textwidth]{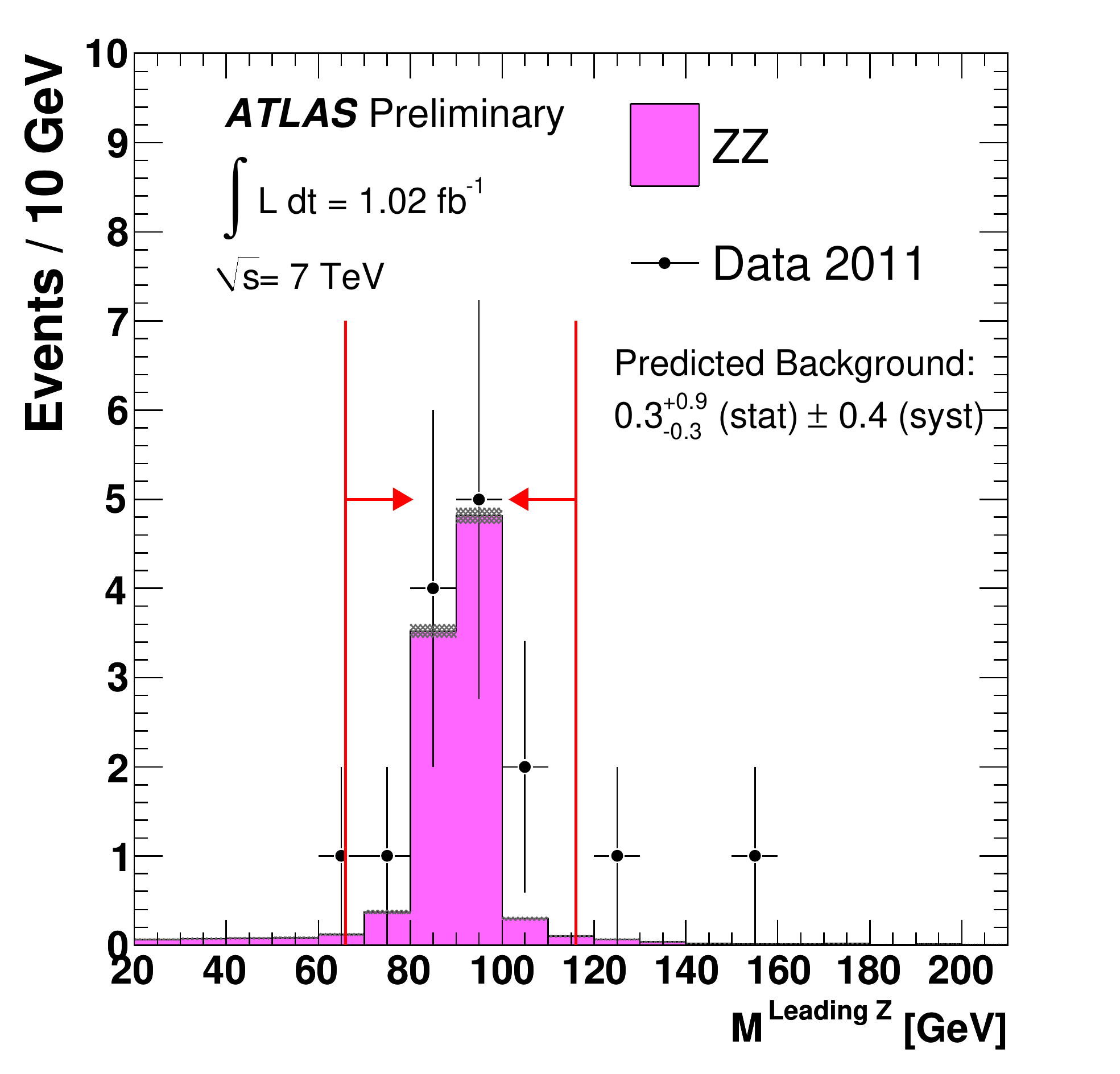}
 \includegraphics[width=0.4\textwidth]{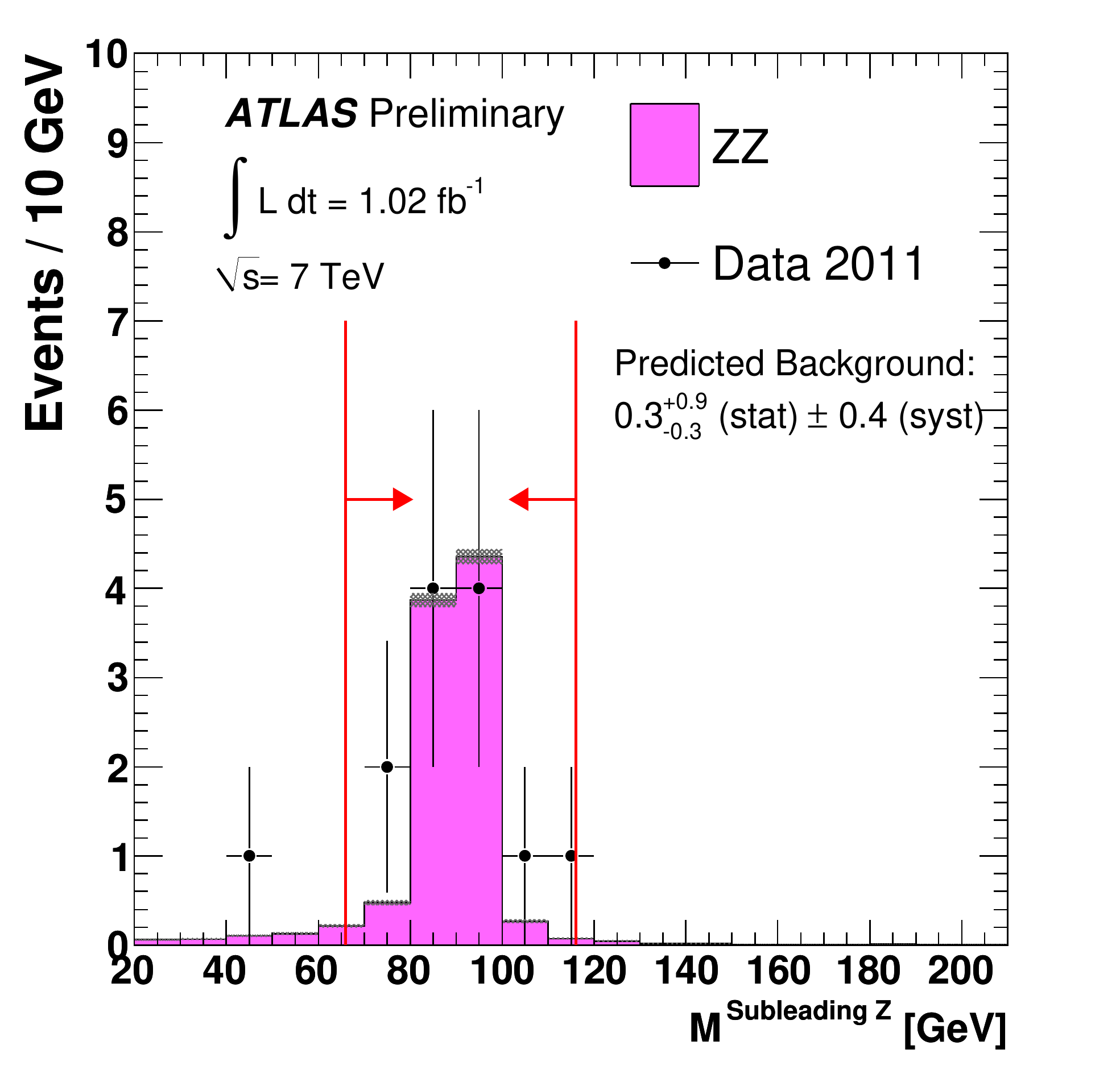}
 \includegraphics[width=0.4\textwidth]{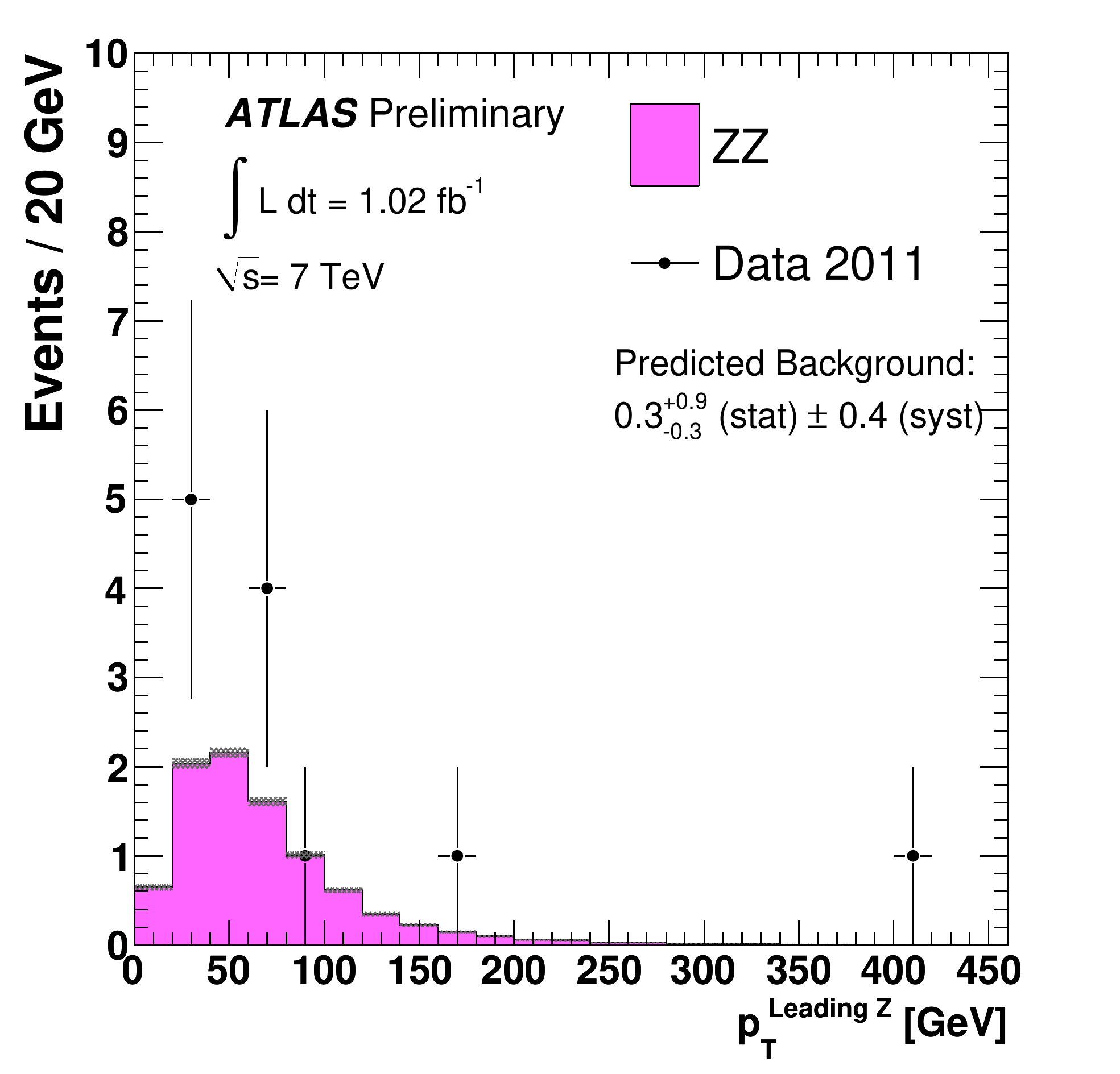}
 \includegraphics[width=0.4\textwidth]{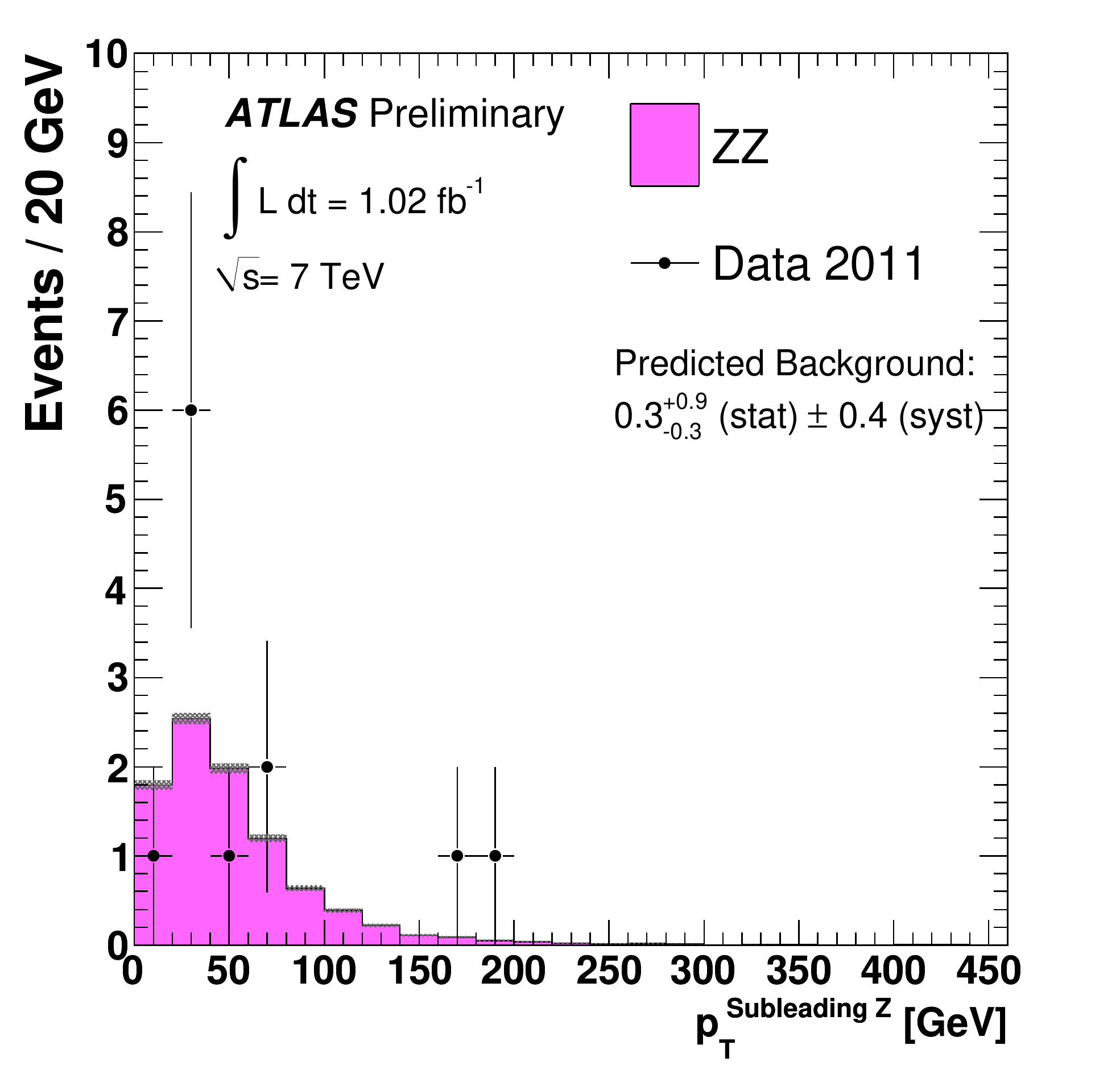}
 \includegraphics[width=0.4\textwidth]{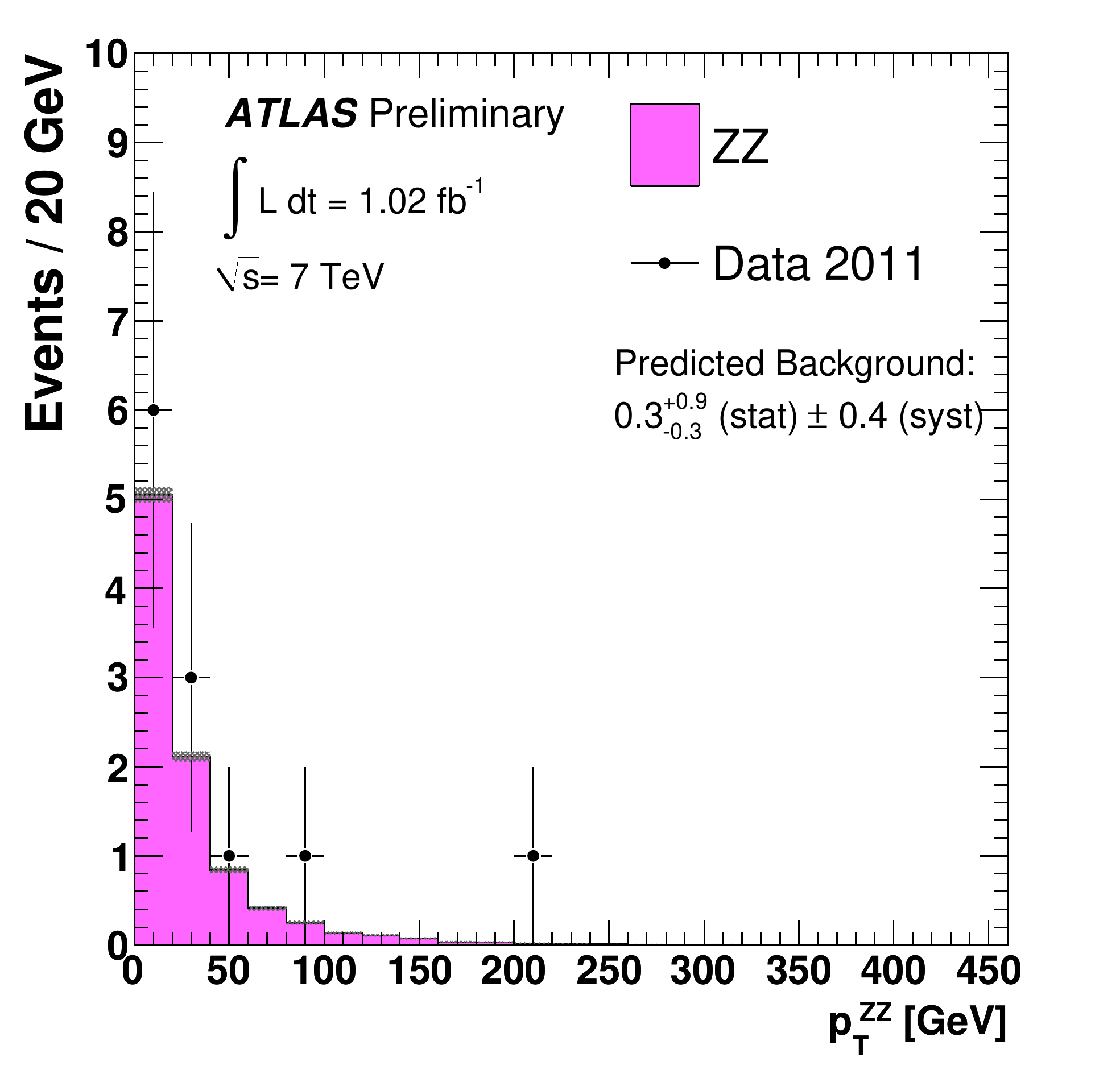}
 \includegraphics[width=0.4\textwidth]{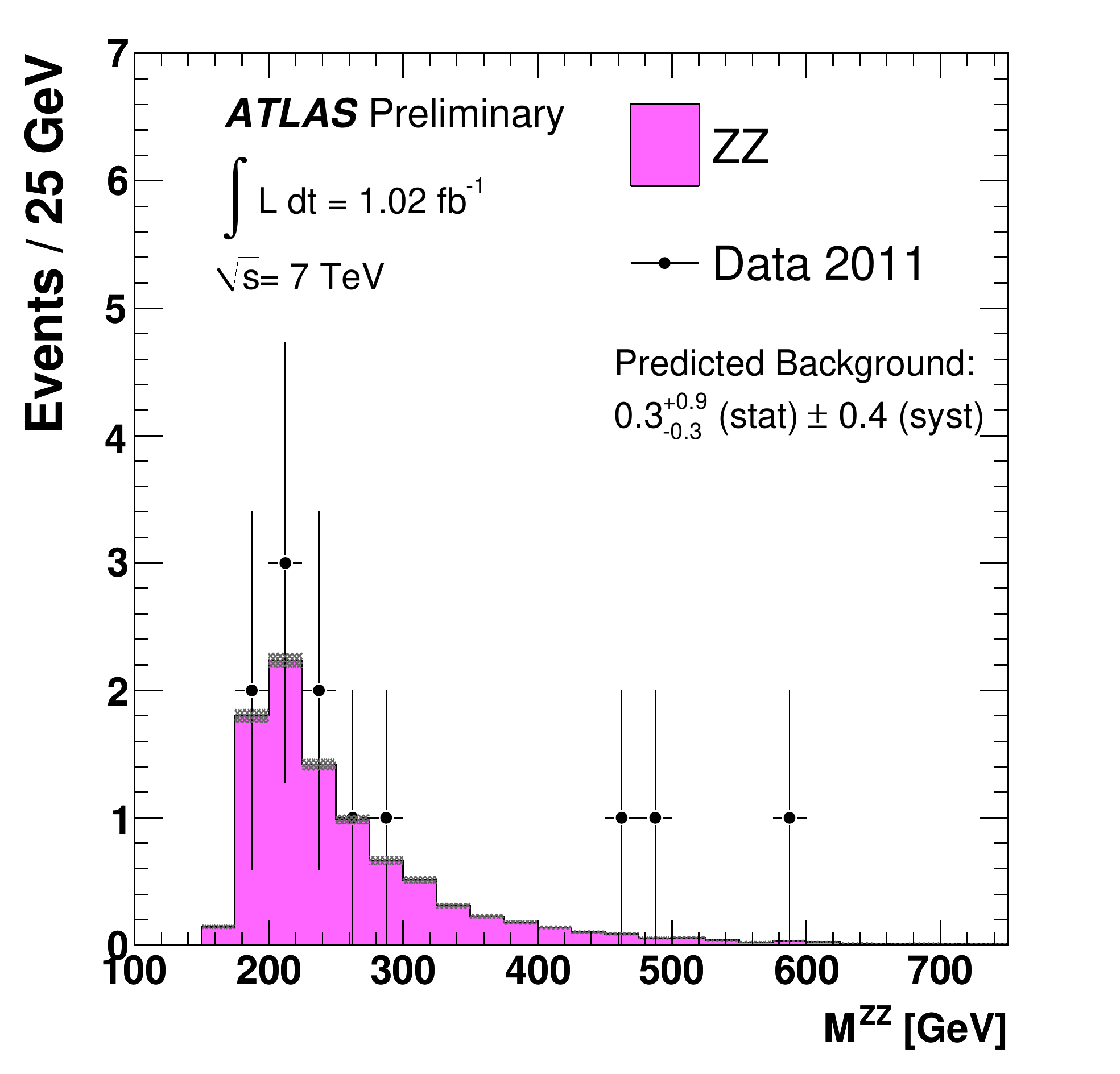}
 \caption{\small Kinematic distributions for \ZZ candidates summed over all four-lepton channels.
          The top row shows the invariant mass of the leading (top left, higher lepton pair $\pt$) and subleading (top 
          right, lower lepton pair $\pt$) lepton pair, after all cuts except on the variable plotted. The red bands 
          indicate the signal region defined by the mass cut.
          The middle row shows the transverse momentum of the leading (middle left) and subleading 
          (middle right) lepton pair, and the bottom row the transverse momentum $\pT^{\ZZ}$ (bottom left) 
          and invariant mass $M^{\ZZ}$ (bottom right) of the four-lepton system, for selected candidates.
          The points represent the observed data and the 
          histogram shows the signal prediction from simulation. The shaded band on the histogram
          shows the combined statistical and systematic uncertainty on the signal prediction. 
          The predicted number of background events from the data-driven background estimate is 
          indicated on the plot. 
   }
 \label{fig:kindists}
 \end{center}
 \end{figure}
\newpage

\section{Limits on anomalous TGCs}\label{sec:TGCLimits}
\begin{table}[htbp]
  \centering
  \begin{tabular}{lcccc} 
    \hline\hline
     Coupling 95\% CI  & $f_4^{\gamma}$ & $f_4^{Z}$ & $f_5^{\gamma}$ & $f_5^{Z}$\\
    \midrule
   $\Lambda=2$ \TeV  & $[-0.15, 0.15]$  & $[-0.12, 0.12]$ & $[-0.15,0.15]$ & $[-0.13, 0.13]$ \\    \hline
   $\Lambda=\infty$  & $[-0.08, 0.08]$  & $[-0.07, 0.07]$ & $[-0.08,0.08]$ & $[-0.07, 0.07]$ \\   \hline

  \end{tabular}
\caption{One dimensional 95\%\ confidence intervals for anomalous neutral gauge boson couplings, where the 
	 limit for each coupling assumes the other couplings fixed at their Standard Model value. A form 
	 factor scale of $\Lambda = 2$ \TeV\ and $\infty$ are both presented. Limits were derived using both statistical
	 and systematic uncertainties; the statistical uncertainties are dominant. 
	}
  \label{ta:TGCLimits}
\end{table}
Limits on anomalous nTGCs are determined using the \ZZ cross section.
The cross section dependency on couplings is parametrized using
fully simulated SHERPA~\cite{Gleisberg:2008ta} events subsequently reweighted using the leading order matrix
element~\cite{Baur:2000ae} within the framework of Bella~\cite{Bella:2008wc}
to account for the multidimensional dependence on acceptance and efficiencies. The reweighting procedure
uses simulated samples with Standard Model as well as non-Standard Model coupling values to remove potential problems from large weights.
One dimensional 95\%\ confidence intervals on the anomalous nTGCs were determined using a maximum profile
likelihood fit to the observed number of events. The systematic errors were included as nuisance
parameters.
The resulting limits for each coupling, determined assuming the other couplings fixed at their Standard
Model value, are listed in Table~\ref{ta:TGCLimits}.
The present limits are dominated by statistical
uncertainties: limits derived using statistical uncertainties alone differ from those in
Table~\ref{ta:TGCLimits} by less than 0.01.
As shown in Figure~\ref{fig:combined},
these limits are comparable with, or are more stringent than, those derived from measurements at
LEP~\cite{bib:LEPEW2006} and the Tevatron~\cite{bib:D0_ZZ1}, although it should be noted that
limits from LEP do not use a form factor, and those from the Tevatron use $\Lambda = 1.2\TeV$.

\begin{figure}[htb]
\begin{center}
\includegraphics[width=.4\textwidth]{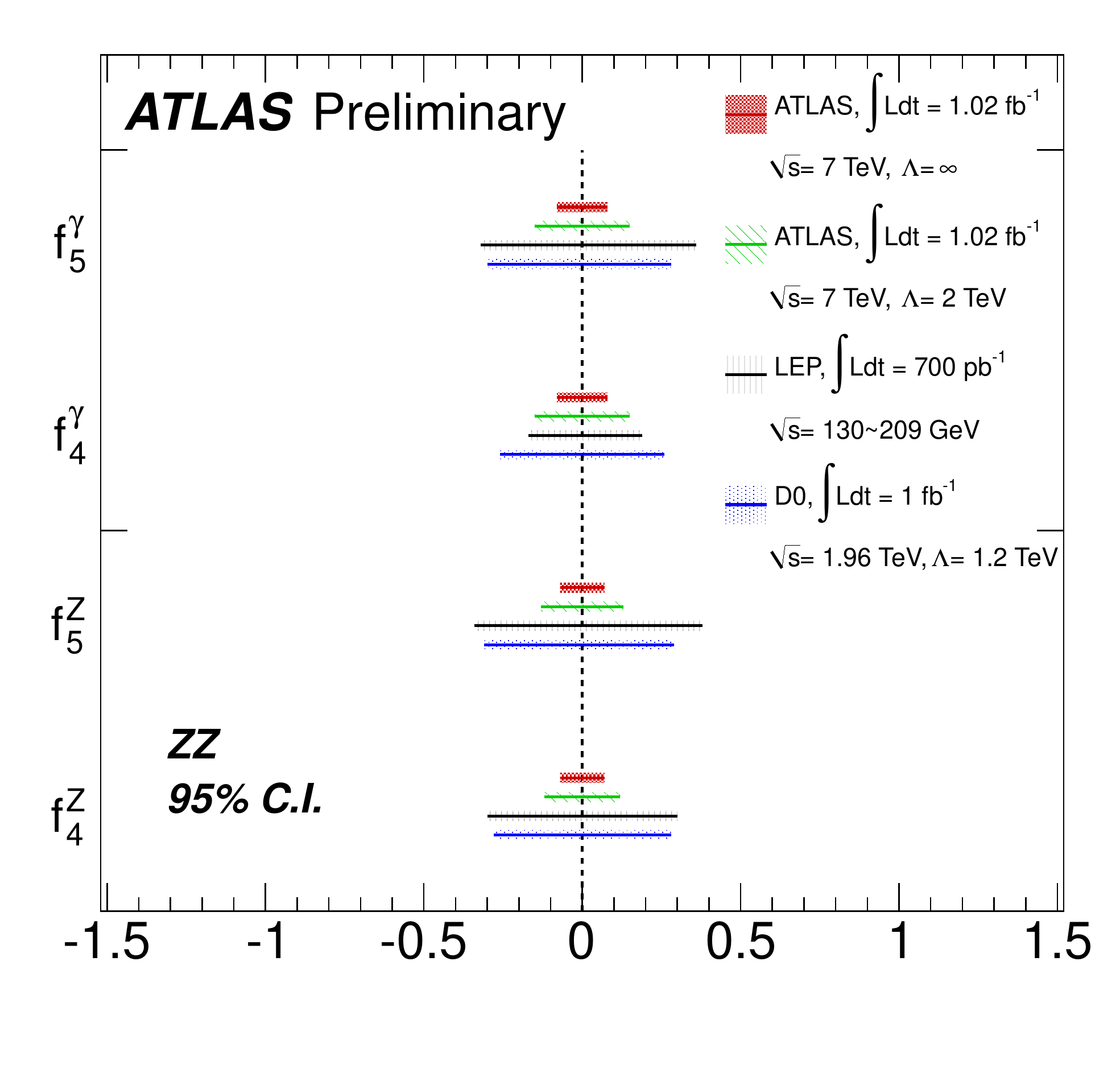}
\caption{\small Anomalous nTGC 95\% confidence intervals from ATLAS, LEP~\cite{bib:LEPEW2006} and 
         Tevatron~\cite{bib:D0_ZZ1} experiments. Luminosities, centre-of-mass energy and cut-off 
         $\Lambda$ for each experiment are shown.}
\label{fig:combined}
\end{center}
\end{figure} 

\section{Conclusion}\label{sec:Conclusion}
The first measurement of the \ZZ production cross-section in LHC proton-proton collisions at 
$\sqrt{s}$ = 7 TeV has been performed by the ATLAS detector, using electrons and muons in the 
final state. 
In a dataset with an integrated luminosity of 1.02 \ifb\  
a total of 12 candidates was observed with a background 
expectation of 
$0.3^{+0.9}_{-0.3}(\mbox{stat}) ^{+0.4}_{-0.3}$(sys).
The Standard Model expectation for the number of signal
events is $9.1\pm0.1$(stat)$\pm0.3$(sys).
The fiducial
and total cross sections were determined to be 

\begin{eqnarray*}
\sigma_{ZZ\to \ll\ll}^\mathrm{fid} &=& 19^{+6}_{-5}\textrm{ (stat) } ^{+1}_{-2}\textrm{ (syst) }\pm 1 \textrm{ (lumi) }\mathrm{fb}\\
\sigma_{ZZ}^\mathrm{tot} &=& 8.4^{+2.7}_{-2.3}\textrm{ (stat) }^{+0.4}_{-0.7}\textrm{ (syst) }\pm 0.3\textrm{ (lumi) }\mathrm{pb}.
\end{eqnarray*}
The result is statistically consistent with the NLO Standard Model total cross section for this 
process of $6.5^{+0.3}_{-0.2}\ \pb$.
95\% confidence intervals for anomalous neutral triple gauge boson couplings are derived which are
compatible with zero, the value they have in the Standard Model. 
These limits are comparable with, or are more stringent than, those derived from measurements at 
LEP~\cite{bib:LEPEW2006} and the Tevatron~\cite{bib:D0_ZZ1}.





\bibliography{ZZDPF2011}

\providecommand{\href}[2]{#2}\begingroup\raggedright\begin{thebibliography}{10}

\bibitem{ATLAS-lumi2011}
{ATLAS Collaboration} ATLAS-CONF-2011-116  (2011)  .

\bibitem{zzres1}
K.~Agashe et al. Phys.\ Rev.\ D  76 (2007)  036006.

\bibitem{zzres2}
L.~Fitzpatrick et al. J. High Energy Phys.  0709 (2007)  013.

\bibitem{Gounaris:2000dn}
G.~J. Gounaris, J.~Layssac, and F.~M. Renard
\href{http://dx.doi.org/10.1103/PhysRevD.62.073012}{Phys. Rev.  D62 (2000)
  073012}.

\bibitem{Ellison:1998}
J.~Ellison and J.~Wudka
  \href{http://dx.doi.org/10.1146/annurev.nucl.48.1.33}{Annu. Rev. Nucl. Part.
  Sci.  48 (1998)  33}.

\bibitem{Baur:2000ae}
U.~Baur and D.~L. Rainwater
\href{http://dx.doi.org/10.1103/PhysRevD.62.113011}{Phys. Rev.  D62 (2000)
  113011}.

\bibitem{Barate:1999jj}
{ALEPH} Collaboration, R.~Barate et al.
\href{http://dx.doi.org/10.1016/S0370-2693(99)01288-5}{Phys. Lett.  B469 (1999)
   287}.

\bibitem{Abdallah:2003dv}
{DELPHI} Collaboration, J.~Abdallah et al.
\href{http://dx.doi.org/10.1140/epjc/s2003-01287-0}{Eur. Phys. J.  C30 (2003)
  447}.

\bibitem{Acciarri:1999ug}
{L3} Collaboration, M.~Acciarri et al.
\href{http://dx.doi.org/10.1016/S0370-2693(99)01065-5}{Phys. Lett.  B465 (1999)
   363}.

\bibitem{Abbiendi:2003va}
{OPAL} Collaboration, G.~Abbiendi et al.
\href{http://dx.doi.org/10.1140/epjc/s2003-01467-x}{Eur. Phys. J.  C32 (2003)
  303}.

\bibitem{bib:LEPEW2006}
{LEP Collaborations ALEPH, DELPHI, L3, OPAL, and the LEP Electroweak Working
  Group}
\href{http://arxiv.org/abs/hep-ex/0612034}{ arXiv:hep-ex/0612034}.

\bibitem{bib:CDF_ZZ}
{CDF} Collaboration, T.~Aaltonen et al.
  \href{http://dx.doi.org/10.1103/PhysRevLett.100.201801}{Phys. Rev. Lett.  100
  (2008)  201801}.

\bibitem{bib:D0_ZZ1}
{D0} Collaboration, V.~M. Abazov et al.
\href{http://dx.doi.org/10.1103/PhysRevLett.100.131801}{Phys. Rev. Lett.  100
  (2008)  131801}.

\bibitem{bib:D0_ZZ2}
{D0} Collaboration, V.~M. Abazov et al.
\href{http://dx.doi.org/10.1103/PhysRevLett.101.171803}{Phys. Rev. Lett.  101
  (2008)  171803}.

\bibitem{bib:D0_ZZ3}
{D0} Collaboration, V.~M. Abazov et al.
\href{http://dx.doi.org/10.1103/PhysRevD.84.011103}{Phys. Rev.  D84 (2011)
  011103}.

\bibitem{ATLAS-CONF-2011-107}
{ATLAS Collaboration} ATLAS-CONF-2011-107  (2011)  .

\bibitem{Campbell:2011}
J.~M. Campbell, R.~K. Ellis, and C.~Williams
J. High Energy Phys.  07 (2011)  018.

\bibitem{bib:ATLASDetectorPaper}
{ATLAS} Collaboration, G.~Aad et al.
  \href{http://dx.doi.org/10.1088/1748-0221/3/08/S08003}{JINST  3 (2008)
  S08003}.

\bibitem{pythia}
T.~Sjostrand et al.
\href{http://dx.doi.org/10.1016/S0010-4655(00)00236-8}{Comput. Phys. Commun.
  135 (2001)  238--259}.

\bibitem{bib:MRST2007LOmod}
A.~Sherstnev and R.~S. Thorne
  \href{http://dx.doi.org/10.1140/epjc/s10052-008-0610-x}{Eur. Phys. J.  C55
  (2008)  553}.

\bibitem{bib:MSTW2008}
A.~D. Martin, W.~J. Stirling, R.~S. Thorne, and G.~Watt
\href{http://dx.doi.org/10.1140/epjc/s10052-009-1072-5}{Eur. Phys. J.  C63
  (2009)  189}.

\bibitem{bib:mcatnlo}
{S. Frixione and B.R. Webber} J. High Energy Phys.  0206 (2002)  029.

\bibitem{madgraph}
J.~Alwall et al.
J. High Energy Phys.  09 (2007)  028.

\bibitem{alpgen}
M.~L. Mangano, M.~Moretti, F.~Piccinini, R.~Pittau, and A.~Polosa J. High
  Energy Phys.  0307 (2003)  .

\bibitem{pythiab}
S.~P. Baranov and M.~Smizanska
\href{http://dx.doi.org/10.1103/PhysRevD.62.014012}{Phys. Rev.  D62 (2000)
  014012}.

\bibitem{bib:ATLASMCpaper}
{ATLAS} Collaboration, G.~Aad et al.
  \href{http://dx.doi.org/10.1140/epjc/s10052-010-1429-9}{Eur.Phys.J.  C70
  (2010)  823--874}.

\bibitem{bib:geant4a}
{S. Agostinelli et al.}
  \href{http://dx.doi.org/10.1016/S0168-9002(03)01368-8}{Nucl. Instrum. Meth.
  A 506 (2003)  250}.

\bibitem{bib:ATLAS_W_Z}
{ATLAS} Collaboration, G.~Aad et al. J. High Energy Phys.  1012 (2010)  060.

\bibitem{ATLAS-CONF-2011-063}
{ATLAS Collaboration} ATLAS-CONF-2011-063  (2011)  .

\bibitem{Gleisberg:2008ta}
T.~Gleisberg, S.~H{\"o}che, F.~Krauss, M.~Sch\"{o}nherr, S.~Schumann,
  F.~Siegert, and J.~Winter J. High Energy Phys.  02 (2009)  007.

\bibitem{Bella:2008wc}
G.~Bella
\href{http://arxiv.org/abs/0803.3307}{ arXiv:0803.3307}.

\end{thebibliography}\endgroup

\end{document}